\documentclass[aps,twocolumn,amsmath,amssymb,fleqn]{revtex4}

\usepackage{amssymb}
\usepackage{amsmath}
\usepackage{amscd}
\usepackage{latexsym}
\usepackage{epsfig}
\usepackage{graphicx}
\usepackage{bbm}

\begin{document}
\title{ Spin Decoherence from Hamiltonian dynamics in Quantum Dots}
\author{D. D. Bhaktavatsala Rao, V. Ravishankar, V. Subrahmanyam}
\affiliation{Department of Physics, Indian Institute of Technology, Kanpur-208016,
INDIA}

\begin{abstract}
The dynamics of a spin-$1/2$ particle coupled
to a nuclear spin bath through an isotropic Heisenberg interaction is studied,
as a model for the spin decoherence in quantum dots. The time-dependent
polarization of the central spin is calculated as a function of the
bath-spin distribution and the polarizations of the initial
bath state.
For short times, the polarization of the central spin shows
a gaussian decay, and at later times it revives displaying nonmonotonic time
dependence. The decoherence time scale depends on moments of the
bath-spin distribuition, and also on the polarization strengths in various
bath-spin channels. The bath polarizations have a tendency to increase
the decoherence time scale.
The effective dynamics of the central spin polarization is shown to be 
described by a master equation with non-markovian features.

\end{abstract}
                      
\maketitle          
                                  
\section{Introduction}

Decoherence is ubiquitous in quantum systems, either because of the interaction 
with the environment \cite{zur}, or because of interventions from the measuring 
apparatus \cite{srin}. An understanding of the phenomenon of decoherence is, 
therefore, essential --  from the view point of the foundational issues as 
well as the dynamics of open quantum systems. In this context, $N$-level
systems are particularly interesting; the state is defined  in a finite
dimensional 
Hilbert space
( the associated phase space is compact), which makes the theoretical analysis
simpler. 
More importantly, there is an abundance of N-level systems as
encountered in 
NMR systems \cite{chuang}, NQR systems, and  polarized photons. 
In addition, one also has
the so called pseudo $N$-level systems - occurring in atomic, semiconductor and
quantum 
dot environments. Although the states of the latter systems are  defined in
infinite 
dimensional Hilbert spaces, energetics (at sufficiently low temperatures)
effectively restrict the states
to lie in a finite dimensional subspace. It is no surprise that  the
dynamics 
of $N$-level
systems (spin systems in short) has been extensively studied.

Decoherence in spin systems has acquired a new importance in view of the rapid
developments 
which are taking place in quantum information theory. Controllable quantum gates
are central 
to quantum computation, and are built of qubits which are physically realized
through 
spin-half (or pseudo spin-half) systems. Proposals for qubits include
the spin of electrons
\cite{priv1,priv2}, nuclear spins \cite{priv3,priv4}and squids
 \cite{priv5,priv6, priv7}. Photons are also potential candidates since highly entangled states 
have been prepared experimentally
by employing the polarization degree of freedom  \cite{zhi}. 
Recently, fullerene-based single-electron transistors \cite{feng} have also been proposed where the electron spin acts as the qubit. Finally, we mention yet another important example involving the spin of the electrons, viz, quantum dot quantum computers (QDQC)\cite{loss2}.
An additional feature of QDQC is that the number of qubits
(electrons) 
can be controlled precisely: starting from zero, electrons can be added one by
one. 
The preparation of quantum gates and their manipulation should be simpler in QDQC.

 Minimising the decoherence involving the electron spin in various environments is, therefore, of great importance. We study in this paper the dynamics of decoherence  with a special focus on
  QDQC. However,  
   our method possesses wider   applicability to instances such as  decoherence in fullerenes, and to  spin systems
where a given reference spin - the central spin - decoheres due to its
interaction with the bath constituted by the other spins.
In QDQC systems, the environment is constituted  by  $GaAs$, $(Ga, Al)As$, and $InAs$. 
The nuclei  $^{69}Ga (\frac{3}{2},~+2.016)$, $^{71}Ga (\frac{3}{2},~+2.562)$, 
$^{27}Al (\frac{5}{2},~+3.6414)$ and $^{115}In (\frac{9}{2},~ +5.534)$ have
 non-vanishing spins and  magnetic moments, as shown in the parentheses.  The magnetic moments are written in units of nucleon Bohr magneton. Since the qubit itself
carries a large magnetic moment, the dominant contribution 
to the decoherence comes from the interaction between the spins. Indeed,  decoherence times have been measured experimentally for qubits both in quantum dot ensembles
   and single quantum dots. These studies find a long decoherence time ($\sim 100ns$) in quantum dot ensembles \cite{Kikk}. Similarly,
 studies on
a single quantum dot \cite{fuj1,fuj2,priv1} indicate that the spin flip rate is
low, implying a long decoherence time again.

Theoretically, there have been a large number of investigations, summarized in a
recent paper by
Schliemann et al \cite{losstopical}. A common feature that underlies these investigations
is the 
assumption of a mean magnetic field $\mathcal{B}$ that is produced by the nuclei
which constitute the environment. This assertion is based on  the observation that
there are $10^4 - 10^6$ nuclei, all with
non-vanishing magnetic moments, that interact with the qubit \cite{erlin,schliemann}.
In this semiclassical approach, it is the time dependence in  $\mathcal{B}$ that
causes the decoherence in the spin state.
Reliable estimates for the decoherence rates have been obtained both analytically  and numerically.

This paper revisits the problem of decoherence with a rather different perspective. We show
that given the phenomenology of QDQC, it is possible to map the hyperfine
interaction between the qubit and the nuclei to 
an isotropic  Heisenberg model with a global interaction between the qubit
spin and the total
nuclear spin. More precisely,  the  qubit interacts with a spin bath, whose
initial state is described by a density matrix
$
\rho_{B} = \sum\lambda_{I_B} \rho_{I_B}.
$
where the coefficients $\lambda_{I_B}$ are  the weights for
the bath to be in a state with a total spin ${I_B}$. Within each spin sector of the bath, the
hyperfine interaction between the bath spin and the qubit spin
 gets recast into an isotropic Heisenberg interaction.
With this simplification, the dynamics becomes exactly solvable: 
the state of the qubit+bath can be determined in a closed form
explicitly. A partial trace yields
 the reduced density matrix of the qubit. The decoherence rate is easily inferred thereafter.

In the next section we obtain an effective Hamiltonian from the basic hyperfine 
interaction (for QDQC), and elaborate on the model in some detail.
We then illustrate the dynamics in the simplest of the cases,
{\it viz}, of two qubits in section III. This example is of practical
importance, exhibiting the so called interaction induced entanglement. After the warm up, the general problem of a qubit interacting 
with a spin bath (as described above) will be addressed. The expression for the
reduced density matrix and the dynamics of decoherence will be investigated
at length. Section IV is devoted to examples and special cases.
Section V shows how the results
can be extended to more general local Heisenberg interactions perturbatively,
at small times. 
In Section VI, we shall set up a master equation satisfied by the reduced
density matrix. 
The non-markovian nature of the evolution, the time-dependent decoherence
rates, and the unitary part of the evolution will be clearly identified.

\section{The Model}
A dominant interaction in QDQC between the electron
spin and the nuclei
is given by the hyperfine coupling of the electron spin to the
nuclear spins:
\begin{equation}
\label{hfh}
{H}_{hf}=\sum_i K_i \vec{S} \cdot { \vec{I}_i}
\end{equation}
where $K_i$ is the interaction strength between the qubit spin $\vec S$
and the $i$'th nuclear spin $\vec I_i$. The coupling 
constant $K_i$ depends on the basic coupling strength and the wave
function of the electron, 
$K_i = K \vert \Psi(\vec{r}) \vert^2_{\vec{r}=\vec{r}_i}$,
where the electron density is evaluated at the location of the nucleus. 
Merkulov et al\cite{merk} have pointed out that the spin orbit interaction,
which is otherwise dominant in spin decoherence, is suppressed here because
the qubit 
is strongly localized. For the same reason the contribution of the phonon mediated 
interactions is also suppressed  at low temperatures \cite{merk}. Thus the
hyperfine interaction between the electron and the nuclei emerges as the
dominant mechanism for spin relaxation.  

In addition to the hyperfine interaction, the electron spin may also be
subjected to an external 
magnetic field which is responsible for its polarization.
Alternatively, a spin polarized electron can be directly
injected into a quantum dot, as   Cortez et al\cite{cortez} have done. Remarkably,
they have also succeeded in `writing' and then subsequently `reading' a quantum
dot. We shall be interested in this ``zero field" scenario. In either case,
the nuclear coupling to the field is relatively negligible because of the
smaller values of the nuclear magnetic moments.  Finally, mention should be made of the 
internuclear dipolar coupling. This coupling (strength is of order
10$^{-12}$eV)  is a small perturbatio to
the other interaction terms. But it has an important role in determining
the initial spin state of the nuclei, and its effect on the subsequent
dynamics is negligible\cite{loss2}.

As mentioned earlier, each electron is in an environment of $10^4-10^6$ nuclei. 
It has been assumed, therefore, that the cumulative effect of the nuclear spins
is to produce an effective  magnetic field $\mathcal{B}$ - the Overhauser field - at the site of the qubit. 
 Likewise, the action of the electron spin on the nuclei
makes $\mathcal{B}$ time dependent, which in turn causes spin flip transitions in the qubit \cite{erlin,losstopical, merk}.
 For an explicit determination of the transition amplitudes,  computational schemes in this 
semi-classical approach involve either taking a gaussian distribution for 
$\mathcal{B}$ \cite{erlin} or  considering special configurations for the
initial state\cite{khaetski,khaprb}.  
The contribution of intra-nuclear interaction to the time-dependent magnetic
field, and the resulting spectral diffusion of the qubit polarization were
considered by de Sousa and Das Sarma\cite{desousa}, and Deng and Hu\cite{deng}.
Let us summarize the semiclassical results in brief. Of the two decoherence times, 
it is is found that  
the phase decoherence time is larger; standard spin flip transition calculations
(using the Fermi golden rule) yield decay profiles which are nonexponential 
\cite{khaprb}. More interestingly, the same authors claim that for a 
large Zeeman field, the polarization  of the electron decays 
$\sim t^{-\frac{3}{2}}$. This surprising result is obtained for a special choice for the
initial state 
of the qubit-nuclear system. Numerical simulations by Schliemann et al
\cite{schliemann}
for a variety of nuclear spin states
have yielded 
several interesting results: (i) the polarization of the electron
decreases in magnitude, with accompanying oscillations whose period is estimated
to be $4 \pi/K$, with only a weak dependence on the initial state; 
(ii) if the nuclei are initially only randomly correlated, the spin dynamics 
of the electron is significantly different in that the decay is much faster. 
A master equation for the qubit density matrix was set up, and the dynamics
was investigated perturbatively by Coish and Loss\cite{coish}, where the
effect of the electronic wave function on the qubit decoherence was also
considered.

Interesting that these results are, the  semiclassical approximation needs
a validation from a more
rigorous analysis. Indeed, the hyperfine Hamiltonian is itself not easily
amenable to a 
full quantum  treatment.  We nevertheless argue that the phenomenology of
QDQC
allows us to map the problem to the more tractable isotropic Heisenberg
interaction. To that end, 
consider a circular geometry for the quantum dot containing the electron
\cite{schliemann}. 
Taking the confinement to be either  parabolic or coulombic, the electron wave
function
$\psi (\vec r)$ is either a 
gaussian or  an exponential with the distance.
The effective coupling to the nuclear 
spins which are farther is accordingly suppressed, with the maximum contribution coming from the
nearest nuclei, all of which are roughly  equidistant from the electron. 
Indeed, in GaAs there are 
about 45 nuclei in a volume of 1 nm$^3$, and the electron wave function is
roughly uniform over a distance of 2-3 $nm$ (while the size of the quantum 
dot is about 20 $nm$)\cite{losstopical,lee}. This translates into about a few
hundred nuclei that interact with the qubit with the same coupling strength 
$K_i$ in Eq.\ref{hfh}, and thus the Hamiltonian
assumes a simple form
\begin{equation}
H_{eff} =  K \vec{S} \cdot \vec{I}_{{B}},
\end{equation}
where $\vec{I_B}=\sum \vec I_i$ is the total spin of the (nearest) nuclei.
 If we were to consider the nuclei in the next nearest circle (next to
nearest 
neighbors in these two dimensional samples),
the Hamiltonian acquires additional terms with another effective coupling  constant  $K^{\prime}$ whose value is exponentially suppressed relative to $K$, and may be treated as a
(small) perturbation. We shall ignore all such higher order contributions here. 
The dynamics has been effectively mapped, to an excellent approximation,
to an isotropic Heisenberg
interaction between the qubit spin and the total nuclear bath spin.

The above Hamiltonian can have a more general role: consider an assembly of $N$
identical spin-half particles interacting  through a Heisenberg interaction.
We can write the Hamiltonian of the spin system as
\begin{equation}
\label{hamgen}
H_{spin} = \sum_{i,j}K_{ij} \vec{S}_i \cdot {\vec{S}}_j,
\end{equation}
where $K_{ij}$ is the interaction strength of a pair of spins $\vec S_i$ and
$\vec S_j$.
For any chosen spin,  denote it by $\vec S$, the other spins would
constitute a bath. We can split the above into terms involving the interaction
of the spin and the bath, and an intra-bath term, as
\begin{equation}
H_{spin}= K_1  \vec S\cdot \vec I_{B_1} + K_2 \vec S\cdot \vec I_{B_2}+
.. ~~H_{bath},
\end{equation}
where we displayed the interaction strength $K_1$ ($K_2$) of the central spin
with
the total spin of the first (second) neighbours $\vec I_{B_1} (\vec I_{B_2})$.
And similarly there are interactions with further neighbours. The last term is the intra-bath intreaction. In these spin systems typical interaction strengths are
$K_1\sim 0.1$eV and $K_2 <K_1$. The number of nearest neighbours is about 6,
depending on the lattice structure. 
Keeping the dominant intreaction (the first term) of the qubit with the bath,
the problem reduces to the effective Hamiltonian given in Eq.2.
Consequently, our foregoing analysis for the qubit dynamics is applicable to
these spin systems as well. The dynamics
in these spin systems has been studied at length in the context of
entanglement generation and propagation\cite{subrarul}. 

We will investigate in Sections III and IV the qubit dynamics exactly for the
model Hamiltonian given in Eq.2, as a function of the initial nuclear bath spin
distribution and nuclear polarization strengths in various spin sectors. 
As we will see later, through the time evolution, the qubit
polarization displays a decoherence
regime (characterized by a gaussain decay with a decoherence time scale $\tau$)
for short times. The decoherence time depends on the initial bath state.
For longer times, the qubit polarization revives, as a
consequence of the quantum coherent evolution of the total system of the qubit
and the bath. The revival time (Poincare time scale) for the subsystem is 
large in comparison with $\tau$. However, for times larger than $\tau$ the
sub-dominant interactions, which were ignored in the simplified model
Hamiltonian, will come into play. The revival time becomes larger when the next
to leading interaction is considered, as the revival time scale is set by the
smallest interaction strength. Thus the revival time in a model Hamiltonian 
dynamics has no significance for the real system, as the weaker interactions
are ignored in the model, which will push the revival time further. 
The decoherence regime boundary, determined by the dominant interactions
contained in the model Hamiltonian, is not affected too much by the inclusion 
of the sub-dominant interactions. As we will see in Section IV, the decoherence time scale from our model Hamiltonian compares very well with the experimental
numbers, implying that our model is an excellent approximation. 
 \begin{figure*}[htb]
   \includegraphics[width=7.6cm]{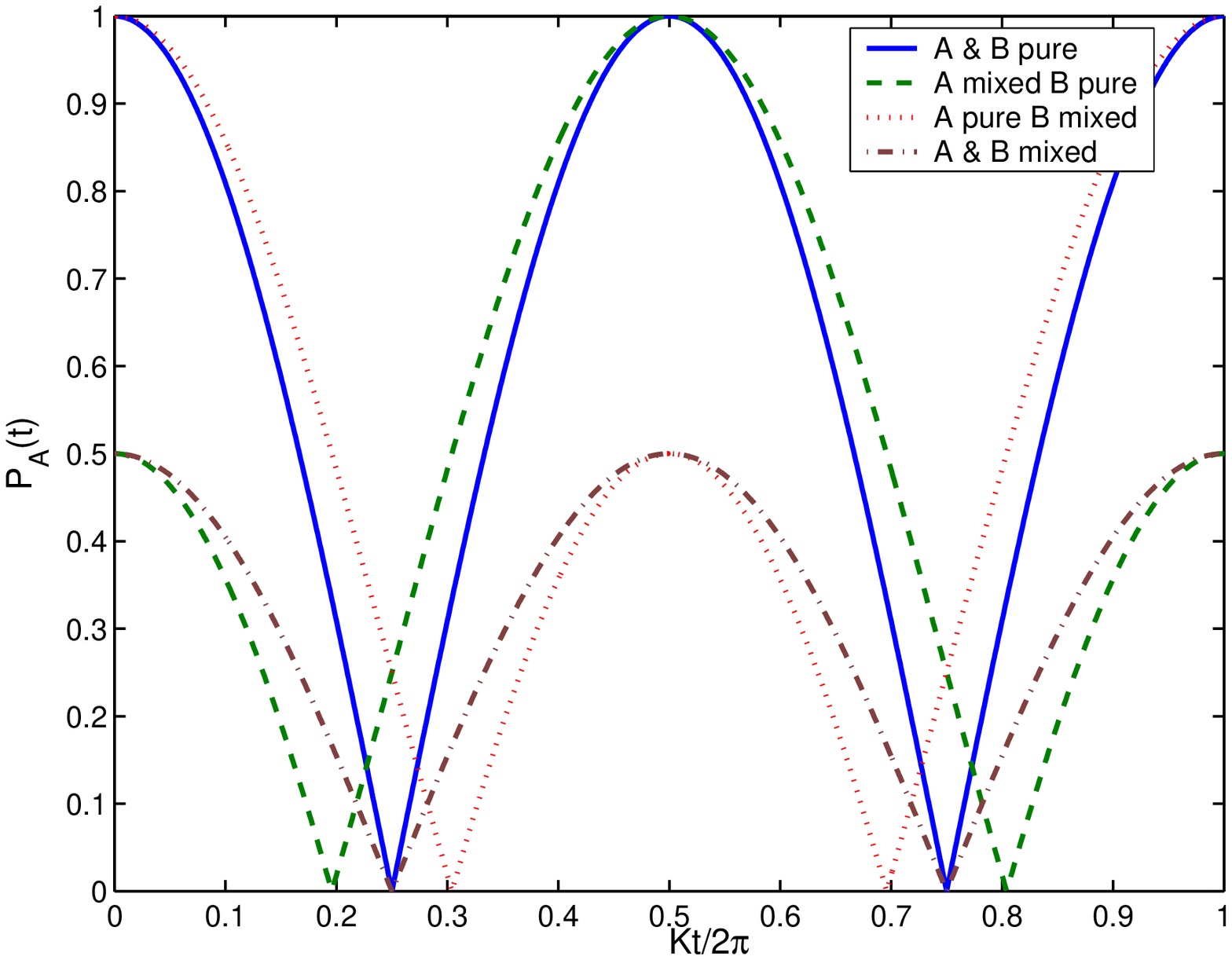}
    \includegraphics[width=7.6cm]{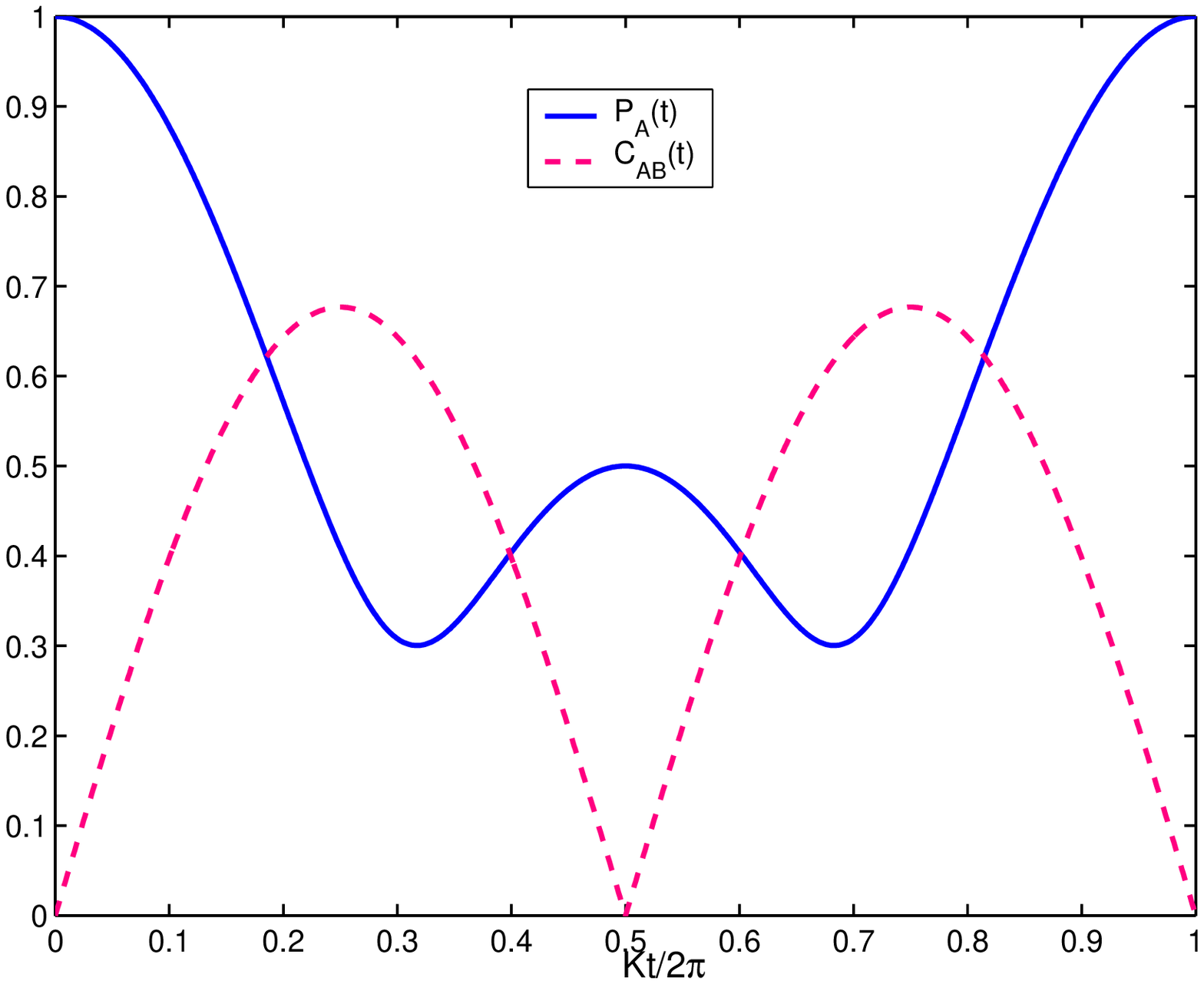}
 \caption{ The polarization $P_A(t)$ of a spin-1/2 particle A
interacting with another spin-$1/2$ particle B is plotted (left figure) for
four different
initial states of the total system A-B.The initial polarizations of A and B are
either $1$ or $0.5$. The $P_A(t)$ and the concurrence measure of
entanglement, $C_{AB}(t)$, between A
and B  is plotted (right figure) with time. The initial polarizations
are given $\vec{P}_A(0)=\hat x$, $\vec{P}_{B}(0)=(-\hat x+\hat z)/2
\sqrt{2}$.}
 \end{figure*}

\section{decoherence with Heisenberg interactions}

\subsection{The initial state}
It is straightforward to specify the initial state of the central spin and the
nuclear spin bath. 
As mentioned earlier, we consider the case where  a prepolarised electron is
injected into the quantum dot. Its state is initially uncorrelated with the
nuclear state. We shall take the state of the total system to be a direct
product of the qubit state and the nuclear bath state.
At $T=0$, the bath will be in its ground state which is sensitive to the
inter-nuclear interactions.
For example, a ferromagnetic interction 
coupling would lead to a fully-polarised bath state. This would give rise
to a large nuclear magnetic field acting on the qubit spin. At nonzero 
temperatures, the bath will be in a mixed state with contributions from many
bath spin sectors with weights for each spin sector $I_B$ (to be fixed by
the energetics and the temperature).
We will consider the case where the intra-bath dynamics conserves the total
bath spin. 
The density matrix of the bath spins may be written, on general considerations
as
$\rho_{B} = \sum \lambda_{{I}_{B}} \rho_{{I}_{B}}$ where the sum is over
all the possible values 
of the total bath spin that the dominantly-interacting nuclei can take. 
In each spin sector the bath can have polarizations of various rank.
The dynamics of the Heisenberg interaction between the qubit and the bath,
does not mix different bath spin sectors. We can determine the dynamics of
the qubit and bath in different spin sectors separately.

A further comment is in order here. Consider the nuclei in a state with a given 
value for the total spin $I_B$. It is customary to consider two extreme
situations:
when the nuclear magnetic field has a nonzero expectation value, and when only
the
fluctuating values survive. In the latter case the bath is usually dubbed as unpolarized.
For a maximally-unpolarised bath, the density matrix is 
$\rho_B = \frac{1}{2^N}\mathcal{I}$. Then,
not only does $\langle{\vec{I}}_{B}\rangle$ vanish, so do all
higher order moments involving the multilinears in the spin operators. By
projection theorem, the expectation values of all the moments of the magnetic
field operator  also vanish. There is no {\it a priori} nuclear magnetic field, and it has to be necessarily induced by the qubit.
On the other hand, should the system be prepared in such a way that 
 $\langle \vec{I}_{B}\rangle =0$, but $\langle {I}_{{B}}^m {I}_{{B}}^n \rangle$ 
is non vanishing, the system is not unpolarized; only the vector polarization
is zero. It is
of course possible that higher rank tensor polarizations may survive. We shall
clarify the contributions coming from  tensor polarizations of various ranks,
i.e., of various higher order magnetic field fluctuations later.

Finally, we shall take all the nuclei to be spin-1/2 particles 
for simplicity. In reality, the nuclei do have higher spins, but the robustness 
of the analysis is unaffected by this approximation.
As a warm-up example, let the bath be  comprised of a single
spin-half nucleus, as in the following.

\subsection{Interaction between two spin-half particles}

Though simple, this case is important because it shows how entanglement is
generated by interactions. 
The Hamiltonian is given by
$H = K \vec{S} \cdot \vec{I}_{B}$. Let the initial state be a direct product state, 
\begin{equation}
\rho(0) = \frac{1}{2}(\mathcal{I}+\vec{P}_A(0) \cdot \vec{\sigma}_A)\otimes 
\frac{1}{2}(\mathcal{I}+\vec{P}_{B}(0) \cdot \vec{\sigma}_{B}),
\end{equation}
where
$\vec{P}_A$ and $\vec{P}_{B}$ denote the initial polarization vectors of
the qubit and the bath respectively. The polarization of the qubit at
subsequent
times, $\vec P_A(t)\equiv Tr \rho(t) \vec \sigma_A$, is easily evaluated to
yield, 

\begin{eqnarray}
\label{onespbat}
\vec{P}_{A} (t)&=&\cos ^2(Kt/2)\vec{P}_{A}(0) + \sin ^2(Kt/2)\vec{P}_{B}(0)
\nonumber \\
&&  +\frac{1}{2}\sin(Kt)\vec{P}_{B}(0)\times\vec{P}_{A}(0).
\end{eqnarray}
The polarization for the nuclear spin B is obtained by interchanging the
labels $A$ and $B$ in the above expression.
Now starting with an initial polarization $P_A(0)=1$, if $P_A(t) <1$ at a
later
time, the it implies a decoherence. However, after decoherence the polarization
(of the particle A) will build up again, as there is an overall periodicity
due to the Hamiltonian dynamics. 

It is easy to see from the above equation that
$\vec{P}_A (t=\pi/K) = \vec{P}_{B} (0)$, and {\it vice versa}, which shows that the 
polarizations get swapped between the qubit and the nucleus at that time.
The decoherence is illustrated in Fig.(1a) where we have plotted 
the variation of $P_A(t)$ with time for four different choices of the initial
state of the total system. Anticipating the analysis in section VI, one may
note that $\vec{P}_{B}$ has a dual role: to cause a precession of $\vec{
P}_A$ (represented by the cross product term in the above equation) and also 
change its magnitude.  It may also be pointed out that since the evolution
of the total system is unitary, there is a revival of the initial state with a 
period $2\pi/K$. The ``decoherence", however, takes place at a much smaller time
scale, determined by the first term in Eq.\ref{onespbat}.
If $\vec{P}_A=\vec{P}_{B}$, the state remains invariant in time.

Now, the initially-polarized qubit (i.e. in a pure state) loses its polarization
(i.e. the state is mixed), due to the interaction. In other words the total
state of the qubit and the nuclear spin picks up entanglement. Through the
time evolution, the entanglement between the two keeps changing
nonmonotonically. The concurrence measure of the entanglement\cite{Wootters}
is easily calculated.
In Fig.1(b) we have plotted the entanglement measure, along with the
polarization $P_A(t)$
as a function of time. From the figure it is clear that  decoherence sets in
(a decrease in the polarization close to zero time) along with the generation
of entanglement, and at latter times the polarization picks up again at the
cost of the entanglement.

The fidelity measure captures how close is $\rho_A(t)$, the reduced density
matrix of the qubit, to its original state $\rho_A(0)$. The average fidelity
at time $t$ is given by
\begin{equation}
\label{defn1}
F_A (t) =  \langle Tr[\rho_A(t) \rho_A(0)]/Tr[\rho^2_A(0)]\rangle,
\end{equation}
where the average is indicated over a distribution of  initial states of the
qubit. We have normalized the fidelity to unity at time $t=0$.
If the averaging is done with a uniform distribution 
over all possible initial pure states, we have 
\begin{equation}
 F_A^{Pure} (t) =  \frac{1}{2}(1 + \cos ^2 (K t/2)). \nonumber
\end{equation}
The average fidelity over all possible initial pure and mixed states is 
obtained as,
\begin{equation}
\label{avefied}
 F _A (t) = 1-3(\frac{\pi}{4}-\frac{2}{3}) \sin^2(Kt/2).
\end{equation}
The decoherence is, of course, not easily apparent in the above expressions.
The fidelity can be less than unity due to either the state is decohered or
the state has changed coherently (a precession of the polarization vector
in an external magnetic field). It may not be possible to separate the
contributions from a coherent unitary evolution and decoherence in the
above equation.

\subsection{Interaction of a qubit with  N spin-half nuclei}

The central problem of the paper will be addressed here. As already mentioned,
we shall take all the nuclei to be spin-half, for the sake of 
simplicity. As argued in section II, the Heisenberg interaction involves, not all the nuclei, but only 
those in the immediate neighborhood of the localized qubit.

The Hamiltonian is then simply given by
\begin{equation}
\label{hamilsolv}
H_{q-b} = K  \vec{S}\cdot \vec{I}_{B}
\end{equation}
The interaction strength $K$ can be written in terms of the hyperfine
parameters; in general it depends on the value of the total spin of the
environment. 

Let the initial qubit-bath state be a direct product,
$
\rho(0) = 
\rho_A(0) \otimes \rho_{B}(0),
$
where the qubit is denoted by $A$, and the bath by $B$.
The initial state of the qubit has the standard form 
$\rho_A(0) = \frac{1}{2}(\mathcal{I} + \mathbf{\sigma}_A \cdot \vec{P}_A(0))$
 where $\vec{P}_A(0) \equiv Tr[\rho_A(0) \vec{\sigma}_A]$.
The initial bath state $\rho_{B}$ can be written in terms of density
matrices corresponding to the various bath-spin sectors, 
$\rho_{B}(0)= \sum \lambda_{I_B} \rho_{I_B}(0)$, which displays the initial bath
state as an incoherent sum of states labelled by bath spin $I_B$, with
weights $\lambda_{I_B}$. 
The nuclear bath state is in general a mixed state in each sector of the
total bath spin. The bath density matrix in a total bath-spin sector $I_B$ 
can be written in terms of tensor polarizations, and spin operators of
various ranks as given by \cite{book}
\begin{equation}
\rho_{{I}_{B}} = \frac{1}{2{I}_{B}+1}[\mathcal{I} + 
{\vec{P}_{{I}_{B}}\cdot \vec{I}_{B}\over I_B} + {3}\sum_{m,n=1}^{3}{{\Pi ^{mn}_
{{I}_{B}} \hat{{Q}}^{mn}_{{I}_{B}}}\over I_B(I_B+1)} + ...],
\end{equation}
here, $\vec P_{I_B}$ is the vector polarization of the bath in the bath-spin
sector $I_B$, and $\Pi_{I_B}^{mn}$ is a component of the rank-two tensor
polarization with the cartesian indices $m$ and $n$. 
In the above expression we have not shown explicitly the terms involving 
higher-rank tensor polarizations.  We shall see later that only these 
polarization terms will be relevant to the time evolution of the qubit
polarization at all times. The components of the second
rank tensor operator are defined by
${{Q}}^{mn}_{{I}_{B}} = 
({{I}}_{B} ^m {{I}}_{B} ^n +{{I}}_{B} ^n {{I}}_{B} ^m)/2 - 
{I}_{B} ^2 \delta_{mn}/3$ \cite{book}. The polarizations $ \vec{P}_{{I}_{B}}
$ and $\Pi_{{I}_{B}}$
are determined by the expectation values of the vector and tensor spin operators
respectively in the $I_B$ sector, $\langle \vec I_B \rangle = \vec P_{{I}_B} ({I}_B+1)/3$ and
$\langle Q_{{I}_B}^{mn} \rangle = \Pi_{{I}_B}^{mn} (2I_B-1)(2{I}_B+3)/10$. With the above definitions
the components of both vector and tensor polarization are of order unity.

The dynamical evolution of the system is governed by the equation $\rho (t) = U \rho (0) U^{\dagger}$, where the time evolution operator
is given by (in a sector with the bath spin $I_B$, and setting $\hbar=1$)
\begin{equation}
U \equiv \text{e}^{-itH_{q-b}} = {\rm e}^{-{iKt\over4}}(a_{I_B}(t) + b_{I_B}(t)\vec{S} \cdot \vec{I_B}).
\end{equation}
The time-depedent coefficients are $a_{I_B}(t) = \cos(\Lambda t)
 - i \sin(\Lambda t)/(2I_B+1)$ and
$b_{I_B}(t) = -4i \sin(\Lambda t)/(2I_B+1),$ where $2\Lambda =
K(I_B+1/2)$.
The Heisenberg Hamiltonian makes the dynamics exactly solvable. The various
components of the total density matrix, corresponding to different bath spin
$I_B$, can be time evolved separately. 
After determining the state at any time $t$, a partial trace over
the bath degrees of freedom yields the expression for the reduced density
matrix $\rho_A(t)$ of the qubit.
We can represent the reduced density matrix as
\begin{equation}
\rho_A(t) = {1\over 2} (1+ \vec P_A(t)\cdot \vec{\sigma}_A).
\end{equation}
The polarization vector $\vec P_A(t)$ carries all the information about the
qubit, $P_A(t)=1$ for a pure state, $P_A(t)<1$ for a mixed state. It
depends on the bath-spin distribution, and the
polarization strengths of the bath in each total spin channel.
The polarization vector of the qubit at any time $t$ is obtained as
\begin{eqnarray}
\label{manyspinpol}
\vec P_A(t) &=& f(t) \vec P_A(0) + \vec g(t) + \vec h(t) \times \vec P_A(0) \nonumber \\
&& + \sum_{m,n} \tilde{\Pi}_{mn}(t) P_A^m(0) \hat e_n.
\end{eqnarray}
In the above the sum is over cartesian components ($m,n=x,y,z$) and $\hat e_n$
stands for a cartesian unit vector.
The time-dependent coefficients depend on the bath-spin distribution, 
and are given by
\begin{eqnarray}
f(t)=& 1-\sum_{{I}_{B}}\lambda_{{I}_{B}}\frac{4{I}_B({I}_B+1)}{3({I}_B+1/2)^2}
\sin^2 [({I}_B+{1\over2}){Kt\over2}],\nonumber\\
g(t)=& \sum_{{I}_{B}}\lambda_{{I}_{B}}\frac{2({I}_B+1)}{3({I}_B+1/2)^2}\sin^2
[({I}_B+{1\over2}){Kt\over2}] \vec{P}_{{I}_{B}},\nonumber\\
h(t)=& \sum_{{I}_{B}}\lambda_{{I}_{B}} \frac{{I}_B+1}{3({I}_B+1/2)} \sin [(I_B
+{1\over2})Kt]
\vec{P}_{{I}_{B}},\nonumber\\
\tilde{\Pi}_{mn}(t)=&\sum_{{I}_{B}}\lambda_{{I}_{B}}\frac{4{I}_B({I}_B+1)-3}{5(
{I}_B+1/2)^2}\sin^2 [(I_B+{1\over2}){Kt\over2}]\Pi^{mn}_{{I}_{B}}.\nonumber\\
\end{eqnarray}
The coefficient $f$ depends only on the spin distribution, and is independent
of bath polarizations.  All the other coefficients depends on various
polarizations that may be present in the initial nuclear bath state.
It is clear from Eq.\ref{manyspinpol} that $\vec{P}_A$  couples utmost
to the second-rank tensor polarization. It is a straightforward
consequence of the Wigner Eckart theorem that
the higher rank tensors do not couple. We conclude that {\it all} nuclear states which have the same values of vector and second rank tensor polarizations have indistinguishable dynamics.
However, the dynamics of a central spin-one particle (its state will be
described by a vector polarization and a rank-2 tensor polarization) will have
contributions from rank-3 and rank-4 tensor polarizations of the bath as well. 

For small times the polarization shows a gaussian decay behaviour (see Fig.2).
This can
be seen by expanding the time-dependent cofficients shown above. We have the
leading-order time dependence,
$f\approx 1-w_f t^2, g\approx K\sqrt{w_h}t^2/2 , h\approx \sqrt{w_h} t ,
\tilde \Pi_{mn}\approx w_{mn} t^2$, where
\begin{equation}
w_f={K^2\over 3} \sum \lambda_{{I}_B} {I}_B ({I}_B+1)
\end{equation}
\begin{equation}
w_h=\{ {K\over 3} \sum \lambda_{{I}_B} ({I}_B+1)\vec P_{{I}_B} \}^2
\end{equation}
\begin{equation}
w_{mn} = {K^2\over 5} \sum \lambda_{{I}_B} ({I}_B ({I}_B+1) -{3\over4})
\Pi_{{I}_B}^{mn}.
\end{equation}
With the above expansion of the time-dependent coeffcients, the small-time
behaviour of the polarization shows a  gaussian decoherence,
\begin{equation}
P_A(t) \approx P_A(0) {\rm e}^{- ({t\over \tau})^2},
\end{equation}
where the gaussian time scale is given by
\begin{equation}
{1\over \tau^2} \equiv w_f -{w_h\over 2} \sin^2\theta_h - {K\sqrt{w_h}\over
2 P_A(0)} \cos{\theta_g} - \sum {P_A^m P_A^n\over P_A(0)^2} w_{mn},
\end{equation}
where the angles appearing above are given by, $\cos \theta_h = \hat h(0)\cdot
\hat P_A(0),\cos \theta_g = \hat g(0)\cdot\hat P_A(0)$. The decoherence time scale
$\tau$
depends on the bath-spin distribution (through the moments), and also the
magnitudes of the vector and tensor polarizations. Nonzero bath polarizations
have
a tendency to increase the time scale over which the gaussian decay takes
place. However, the dominant contribution to decoherence comes from the
first term (i.e., from the function $f(t)$ in Eq.14) which does not depend on
bath polarizations. 

Another perculiar feature is that the time scale also has the initial
qubit state dependence through the appearance of $P_A(0)$ in the above 
expression, which is the hallmark of nonmarkovian dynamics. 
Starting with qubit pure states ($P_A(0)=1$), initially there will be a decay,
i.e.
decoherence, over a time scale $\tau$. At later times, the polarization will
again grow, showing a 
nonmonotonic behaviour, as we will see in the next section in specific examples
of the bath-spin distribution. 
We will also consider more general local interations, unlike the global
interaction considered above, where the qubit can interacting with several
nuclear spins with the different interaction strengths.
We will calculate the gaussian decoherence time scale by a direct expansion of
the time evolution operator in powers of time.

It should be noted that,
for a generic markovian evolution, an exponential decay is expected.
However, the case we have considered where the central spin-1/2
is evolved with the spin bath through Hamiltonian dynamics, 
a gaussian decay is typical as the process of elimination of the bath degrees
of freedom is a nonmarkovian process. 
A comparision of markovian and nonmarkovian dynamics will be done
in the last section, where we construct a master equation that describes
the effective dynamics of the central spin-1/2 particle.

Before we proceed to discuss special cases and examples, we write the
expression for the average fidelity for the qubit, averaging over all
possible initial states of the qubit, 
\begin{equation}
\label{avfied}
\langle F_A (t) \rangle = 1-(\frac{\pi}{4}-\frac{2}{3})\sum_{{I}_{B}} 
\frac{\lambda_{{I}_{B}}{I}_{B}({I}_{B}+1)}{10({I}_{B}+1/2)^2}\sin ^2 ({I}_{B} + 1/2){Kt\over2}
\end{equation}
If the averaging is done over all possible initial pure states only, we have
is confined to the pure states only, 
\begin{equation}
\langle F_A (t) \rangle = 1-\frac{2}{3}\sum_{{I}_{B}} 
\frac{\lambda_{{I}_{B}}{I}_{B}({I}_{B}+1)}{({I}_{B}+1/2)^2}\sin ^2 ({I}_{B} + 1/2){Kt\over2} \nonumber
\end{equation}

\section{Examples and special cases}
\subsection{Unpolarized bath}
For an unpolarized bath, polarizations of all ranks are identically zero {\it i.e.}, 
$\langle \vec{I}_{{B}} \rangle = \langle {I}^m_{{B}} {I}^n_{{B}}\rangle = \langle {I}^m_{{B}} {I}^n_{{B}}...\rangle \equiv 0$. The initial bath state is 
$\rho_{B}(0) = \frac{1}{2^N}\mathcal{I}$.
For this case, the polarisation of the qubit at any time is given by
\begin{eqnarray}
\vec{P}_A(t) = f(t)\vec {P}_A(0)
\end{eqnarray}
where $f(t)$ is given in Eq.14 with 
 $\lambda_{{I}_{B}}=\frac{1}{2^N}C^{N}_{\frac{N}{2}-{I}_{B}}\frac{(2{I}_{B}+1)^2}{{N\over2}+{I}_{B}+1}$. We note that 
the decay time scale is determined by small time behaviour of $f(t)$. Expanding $f(t)$ upto $\large{O} (t^2)$ we get
\begin{equation}
\label{unpoleq}
\vec{P}_A(t) = (1-NK^2t^2/4)\vec{P}_A(0) \approx e^{-t^2/\tau^2}\vec{P}_A(0)
\end{equation}
where the decay time sclae is $\tau = \frac{2}{K\sqrt{N}}$. .
Using the known experimental decay time scales of order 10-100 nanoseconds, and
estimating 
the number of nuclei interacting with the qubit to be, $N\sim 100$ (see
Section II), the typical effective coupling strength relavant for QDQC systems
works out to be, $K\sim 10^{-8}\rm{eV}$. 
These results are in broad agreement with the results obtained by \cite{lee}.
This agreement justifies writing an 
effective Heisenberg interaction of the qubit with the total bath spin.

In Fig.(2) we have plotted $P_A(t)$ for an unpolairzed bath composed of $N$=10, 
30, 50 particles. One can clearly see the $N$ dependence on the
initial decay. As $N$ grows large, the value of $P_A(t)$
falls to 1/3 of its initial value rapidly, and stays there for a long time. 
In general when 
$\rho_{{B}}(0) = \sum \lambda_{
I_B} \mathcal{\hat{I}}_{I_B}/(2I_B+1)$,
the gaussian decay time scale is given by $K\tau =1/  \sqrt{<\hat I_B^2>/3}$.
We can also see  in Fig.2 (inset) the revival of the qubit polarization over
a time scale of the order $\hbar/K$, which is a consequnce of the coherent
quantum
evolution of the total system (the qubit and the bath). However, this revival
time scale is not a physical time scale for the qubit, as we inlcude the
next to leading inteaction strength $K^{\prime}$ (which is exponentially smaller
than the dominant interaction strength $K$, see Section II),
the revival time will become larger $\sim \hbar/K^{\prime}$.
The decoherence time scale shown above, however, will
change only slightly due to the inclusion of the sub-dominant interactions.
Thus though our model Hamiltonian is an excellent approximation for the
decoherence time scale, the revival times in the dynamics are not physical
for the real systems. 
\begin{figure}[htb]
  \begin{center}
    \includegraphics[width=8.0cm]{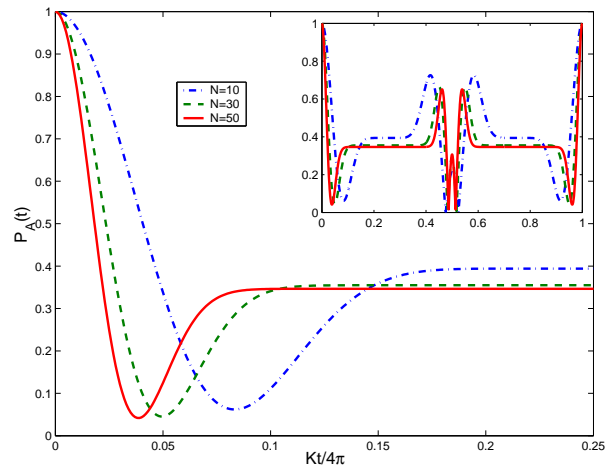}
  \end{center}
\caption{ The polarization $P_A(t)$ of a spin-1/2 particle, interacting with
a completely unpolarized bath
composed of $N$ = 10,30,50 spin-$1/2$ particles, is plotted as a
function of time, over a complete period. The inset shows the short-time
behaviour, a gaussian decay, of the polarization.  }
\end{figure}

In the above examples, the
initial bath state is completely isotropic, and consequently there is no magnetic field produced
by the bath. 
By projection theorem, all the moments of the magnetic field also vanish identically as argued earlier.  We conclude that the
decoherence is a higher order effect in the following sense:   the magnetic field produced by
the qubit polarizes the bath, which in turn produces a magnetic field at the site of the qubit. 
There is no
{\it a priori} fluctuating magnetic field, contrary to the statement found  in
literature. This conclusion holds, of course for a larger class of initial nuclear states whose vector polarization and rank-2 tensor
polarization are absent; we have already remarked that
the higher rank tensors are of no consequence.  To illustrate, let
 $\rho_{B}(0) = \sum_{{I}_{B}}\lambda_{{I}_{B}}
\mathcal{\hat{I}}_{B}/(2{I}_{B}+1)$.
The bath is maximally unpolarized in each spin sector. The polarization of the
bath in a sector $I_B$ can be found easily (by a partial trace over the
qubit degree of freedom), and the total polarization of the bath,
$\vec P_B(t)\equiv \sum \lambda_{I_B} \vec P_{I_B}$, is given by 
\begin{equation}
\vec P_B(t)= \vec P_A(0) \sum \lambda_{I_B} {2{I_B}\over (I_B+1/2)^2}
\sin^2{(I_B+{1\over2}){Kt\over2}} 
\label{bathpol}
\end{equation}
>From the above expression we can see that the induced polarization in each spin sector grows with time; this in turn effects the dynamics of the qubit spin.
The subsequent evolution leads to the decoherence in the spin state. This is
illustrated in Fig.(3), where we have plotted the total induced bath
polarization, $P_B(t)$, for initially-unpolarized nuclear bath with a gaussian
bath-spin distribution .

\begin{figure}[htb]
\begin{center}
\includegraphics[width=8.0cm]{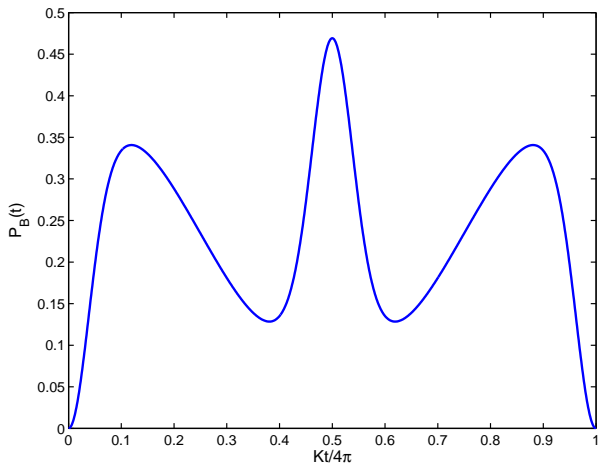}
\end{center}
\caption{ The nuclear bath vector polarization $P_B(t)$ is plotted against time for
one complete period, for the case where a spin-1/2 particle is interacting
with an initially-unpolarized bath composed of $N$=100 spin-1/2 particles,
and a gaussian bath-spin distribution $\lambda_{I_B} \sim 
\exp(-0.1{I_B}^2)$.}
\end{figure}

\subsection{Fully-polarized bath}

We now consider the other extreme, a fully polarized nuclear bath.  All
the nuclear spins are parallel. This example is of relevance
when the inter nuclear coupling is ferromagnetic.
The resultant magnetic field at the site of the qubit can be large, and
may be identified with the Overhauser field.  
 The semiclassical approach may be expected to work the best in
this case.  We shall present below  an exact analysis. 
The initial bath state here is a pure state with bath spin $I_B=N/2$.
$\rho_{B}(0) = \vert \uparrow \uparrow ........\uparrow \rangle \langle \uparrow \uparrow ........\uparrow \vert.$ The vector and tensor polarizations for
this state are 
\begin{equation}
\label{polvals}
\hspace{-3mm} 
\vec{P}_{N\over2}(0) =  \frac{3N}{N+2}\hat{z};{\Pi}_{N\over2}(0) = \frac{5N}{6(N+3)}\left( \begin{array}{ccc} -1 & 0 & 0 \\  0 & -1 & 0 \\ 0 & 0 & 2 \end{array} \right)
\end{equation}
Substituting these values in Eq.\ref{manyspinpol} we get (with $\Lambda=
(N+1)K/4$),
\begin{eqnarray}
P^z_{A}(t) = P^z_{A}(0) +
\frac{2N}{(N+1)^2}\sin^2(\Lambda t) (1-2P^z_{A}(0)) \nonumber \\
\tilde P_{\perp}(t) ={N\over N+1} \left( \begin{array}{cc}
{1\over N}+\cos 2\Lambda t
& -\sin 2\Lambda t\\
\sin 2\Lambda t& 
{1\over N}+\cos 2\Lambda t \end{array}\right ) \tilde P_{\perp}(0)\nonumber
\\ \end{eqnarray}
In the above we have denoted the transverse component of $\vec P_A$ by
a two-component column vector $\tilde P_{\perp}$.
The expressions for the transverse components of the electron spin are the
same as those obtained by placing the electron in a constant magnetic field
$ \vec{\mathcal{B}} =(N+1)K/2 \hat{z}$ apart from terms of order $1/N$.  For
large $N$,
the polarization hardly changes from its initial value. For an inital qubit
pure state with $\vec P_A(0) = -\hat z$, we get from Eq.19, the
gaussian decoherence time scale, $\tau =\sqrt{2} /K\sqrt{N}$. These results
are consistent with the work of Taylor et al\cite{taylor} (however, there is
an apparent discrepancy, which we trace to a  trivial algebraic error in that
paper).

The reason for such a long-lived polarization of the qubit, in the above
example, can be traced to
the energetics of the dynamics. Consider the total state of the system to
be a direct product of the qubit in down spin state and the bath in a fully
polarized state. As the total z-component of the spin ($S^z+I^z$) is conserved
in the evolution, the sate at any later time can be written as
$|{\psi(t)}\rangle=c_1(t)|{\downarrow}\rangle|{N\over2}\rangle + c_2(t)
|{\uparrow}\rangle
|{{N\over2}-1}\rangle$, where $|{N\over2}\rangle$ and $|{{N\over2}-1}\rangle$
denote the
nuclear bath states with $I^z={N\over2}, {N\over2}-1$ respectively. The energy
difference
of this state with the initial state (with $c_2(0)$=0) is given by
$\delta E/K = N(1-|c_1(t)|^2) + \sqrt{N} Re(c_2^{\star}(t) c_1(t)$. Thus, for
large $N$, if the coefficient $c_1(t)$ changes even slightly, there is a large
change in the energy, hence the corresponding transition is not favorable. 
This implies that the qubit polarization cannot change much through the 
evolution. In constrast, 
if we apply a magnetic field of order $KN/\mu_B$ for the qubit along direction
opposite to the nuclear polarization, then the qubit polarization is given by
$P^z_A(t)= \cos (\sqrt{N} Kt) P^z_A(0)$, which shows that the iniial
polarization is not preserved. In fact, there will be coherent oscillations
between the two states shown above. This situation can be used in nuclear
memory where the information coded into the qubit initial state at $t=0$ can
be transfered to the nuclear state through the time evolution. These
considerations were treated by Taylor et al\cite{taylor}.
                                                                                
\begin{figure}[htb]
  \begin{center}
    \includegraphics[width=8.0cm]{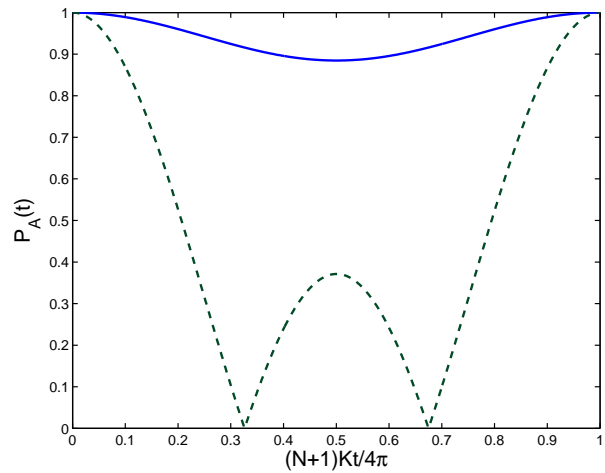}
  \end{center}
\caption{The polarization of a spin-1/2 particle interacting with a nuclear
spin bath with a given total spin, is plotted against time.
The initial polarization is $P_A(0)=-\hat z$. The two different cases shown are
a fully-polarized bath in a pure state with the vector
and tensor polarizations as shown in Eq.25 (solid line), and
a bath in a mixed state with a vector polarization as shown in Eq.25 and
zero tensor polarization (dashed line). A nonzero tensor polarization increases
the initial gaussian decay time scale.}
\end{figure}
In the case of a fully polarized bath, the role played by the
tensor polarization is often glossed over. In fact due to the presence of the
tensor polarizations (see Eq.25), the bath is in a pure state. To highlight
the role played by the tensor polarization in the above case, and show its
importance, 
let us consider the following mixed state $\rho_B(0) = \frac{1}{N+1}(\mathcal{I}+\frac{1}{{{I}_B}} P^z_{{I}_B} {{I}^z_B})$.
For $P^z_{{I}_B} = \frac{3N}{N+2}\delta_{{{I}_{B}},N/2} $ this state has a
vector polarization same as that of the fully-polarized bath that we considered
above, but zero tensor polarization. Again considering an initial qubit pure
state with $\vec P_A(0)=-\hat z$, from Eq.19 we get the decoherence time scale
as $\tau=2\sqrt{3}/K\sqrt{N(N+5)}\sim 1/N.$
In Fig.4 we have plotted $P_A(t)$ when the qubit is interacting with baths of 
same vector polarization but different tensor polarizations.
\begin{figure}[htb]
  \begin{center}
    \includegraphics[width=8.0cm]{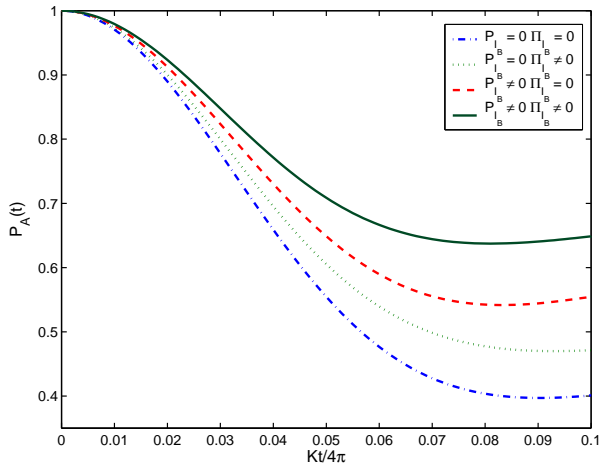}
  \end{center}
\caption{ The polarization $P_A(t)$ of a spin-1/2 particle interacting with a
bath
composed of $N = 100$ spin-$1/2$ particles and the bath spin distribution
$\lambda_{{I}_{B}} \sim \exp(-0.1{I}_{B}^2)$. The
initial polarizations of the bath in various spin sectors are
(i) $\vec{P}_{{I}_{B}} = \hat\Pi_{{I}_{B}} = 0$
(ii) $\vec{P}_{{I}_{B}} = 0$, $\hat\Pi_{{I}_{B}} \ne 0$
(iii) $\vec{P}_{{I}_{B}} \ne 0$, $\hat\Pi_{{I}_{B}} = 0$
(iv)$\vec{P}_{{I}_{B}} \ne 0 \hat\Pi_{{I}_{B}} \ne 0 $.}
\end{figure}
In the absence of
the tensor polarization one can see that $P_A$ decreases rapidly to zero, and
in contrast $P_A$ changes very little in the presence of tensor polarization
of the spin bath. The decoherence time scale, or the gaussian decay time scale,
is much larger when tensor polarization is nonzero.

In Fig.5, we have shown the short-time behaviour of $P_A(t)$, with the
initial qubit polarization
$\vec P_A(0)=(\hat x +\hat z)/\sqrt{2}$ for a few
different bath states, in each
case with a gaussian bath-spin distribution, $\lambda_{I_B}\sim \exp (-0.1
I_B^2)$.
The four cases considered here are (i) no bath polarizations (ii) no
vector polarization but a nonzero tensor polarization  (iii) a nonzero vector
polarization but no tensor polarization (iv) nonzero vector and tensor
polarizations.
The polarizations when nonzero in each bath-spin sector are
given by
\begin{equation}
\vec{P}_{I_B}(0) = \hat{z};{\Pi}_{I_B}(0) = \frac{1}{3}\left( \begin{array}{ccc} -1 & 0 & 0 \\  0 & -1 & 0 \\ 0 & 0 & 2 \end{array} \right).
\end{equation}
It can be seen that the decay of the qubit polarization is slowest
when both vector and tensor polarizations of the bath
are non-zero, and it is fastest when there are no bath polarizations. 

\section{Short-time behaviour with local interactions}

One feature that has emerged from the above is that the decoherence (dephasing
in the case of ensembles of quantum dots) is governed  entirely by the
short-time structure of the evolution operator. 
This suggests that we can use a perturbative solution, to capture the short-time
behaviour of the polarization. Then we may as
well enlarge the class of Hamiltonians to include local interactions and also
include an external magnetic field $\vec{ \mathcal{B}}$. 

Consider the Hamiltonian
\begin{equation}
H = \vec{S} \cdot \sum_{i}^{N} J_i \vec{I}_i +  \vec{\mathcal{B}} \cdot \vec{S}, 
\end{equation}
which is the most general Hamiltonian describing the hyperfine interactions in
quantum dots in the presence of an external magnetic field. The exact dynamics
of the
qubit governed by the above Hamiltonian is difficult to solve as the time 
evolution operator does not acquire a simple form (given in Eq.11) as in the 
case of global isotropic
interaction considered earlier. Taking recourse to a perturbative approach,
we exapnd the time-evolution operator. The qubit polarization can be calculated
up to $\large{O} (t^2)$ as, 
\begin{widetext}
\begin{eqnarray}
\label{smallt}
\vec{P}_A(t) =  \vec{P}_A(0) &-& \frac{t}{2} \lbrace \sum_iJ_i(\vec{P}_A(0)
\times \vec{P}_{B_{i}}(0)) + 2\vec{P}_A(0) \times \vec{\mathcal{B}}\rbrace
\nonumber \\ && +\frac{t^2}{4} \lbrace  {\sum_i {J_i}^2
(\vec{P}_{B_{i}}(0)-\vec{P}_A(0)) + \frac{1}{4}\sum ^\prime_{i\ne j}J_iJ_j(\vec{P}_A(0) \times \vec{P}_{B_{j}}(0))\times\vec{P}_{B_{i}}(0) } 
+2(\vec{P}_A(0) \times {\vec{\mathcal{B}})\times \vec{\mathcal{B}}}
\rbrace. 
\end{eqnarray}
\end{widetext}
In writing the above solution we have taken the initial state of the system to
be a direct product, $\rho(0)=\rho_A(0)\otimes\rho_B(0)$, where $\rho_B(0)=
\rho_{B_1} \otimes \rho_{B_2} \cdots \otimes \rho_{B_N}$. The state of the i'th
nuclear spin is given by $\rho_{B_i} = \frac{1}{2}(\mathcal{I} + \vec{
\sigma}_{
B_i}\cdot \vec{P_{B_i}}(0))$, where its polarization is denoted by $P_{B_i}$.

For an unpolarized bath where all $\vec{P}_{B_{i}}(0) = 0$, the above equation
acquires a simpler form 
\begin{eqnarray}
 \vec{P}_A(t) &=& \vec{P}_A(0)-t\vec{P}_A(0) \times \vec{\mathcal{B}} -\frac{t^2}{4}\sum_i {J_i}^2\vec{P}_A(0) \\ \nonumber
 &&+\frac{t^2}{2}(\vec{P}_A(0) \times \vec{\mathcal{B}})\times \vec{\mathcal{B}} 
 \end{eqnarray}
>From the above equation one can easily see that $P_A(t) = [1-\frac{t^2}{4}\sum_i {J_i}^2]P_A(0)$, indicating a gaussian decay. More interestingly
the decay time scale is independent of the external magnetic field. 

For the case of a fully-polarized bath where all the spins are pointing along
the same quantization axis $\hat{z}$ there is no decoherence if the initial 
qubit polarization also points in the same direction, as the state is an eigen
state of the hamiltonian. On the other hand, if the qubit polarization is given
by
$\vec P_A(0)= -\hat{z}$, and the magnetic field along any arbitary direction, 
the polarization can be simply read off from Eq. \ref{smallt} as
\begin{eqnarray}
\vec{P}_A(t) &=& \vec{P}_A(0) -t\vec{P}_A(0) \times \vec{\mathcal{B}} \nonumber \\
&&+ \frac{t^2}{2}\sum_i {J_i}^2 + \frac{t^2}{2}(\vec{P}_A(0) \times \vec{\mathcal{B}})\times \vec{\mathcal{B}} 
\end{eqnarray}
Here the polarization is $P_A(t) = [1-{t^2\over2}\sum_i {J_
i}^2]\vec{P}_A(0)$,
which is again independent of the external magnetic field. From this we
can conclude that for short times, the gaussian decay of the qubit polarization
is completely determined by the sum of squares of the interaction coupling
strengths $J_i^2$ of the qubit with different nuclear bath spins.

 \section{Master equation for the decoherence of the qubit}

In this section we will derive the master equation obeyed by the qubit,
from the explicit solution for the time-dependent density matrix 
(or the polarization) of the qubit given in Eq.13.
The advantage in setting up a master equation is that it displays the
unitary and nonunitary parts of the qubit evolution, and their
dependence on the initial bath state, viz. the nuclear spin distribution and
the polarization strengths. Though the master equation carries the same
information as the time-dependent polarization calcualted above, it displays
the characteristic effective magnetic field seen by the qubit, and the time
scale for decoherence and polarization procesess explicitly.

We now proceed to recast our results as the solution of a master equation
satisfied by the qubit state. 
  The most general master equation for a two-level system
was given by Gorini et.al., 
\cite{gorini} and Lindblad \cite{lindblad}. Using the notation of Gorini et. al,
the master equation for the qubit density matrix can be written as
\begin{eqnarray}
  \label{eqmo}
 \frac{\partial}{\partial t}\rho_{A} (t)&=& -i[H_c,\rho_{A} (t)] \nonumber \\
&& + \frac{1}{2} \sum_{i,j=1}^{3} \Gamma_{ij} \lbrace [\sigma_i, \rho_{A} (t)
\sigma_j] + 
[\sigma_i\rho_{A} (t), \sigma_j] \rbrace. \nonumber \\
 \end{eqnarray}
where, the first term represents  the unitary component of the evolution. The second term is responsible for non-unitary processes such as decoherence, 
polarization and equilibration. The coefficients $(\Gamma_{ij})$ 
 determine the decay or growth of polarization. 
 
 Since we have an explicit solution for the time dependence of the polarization,
the various terms in the above equation can be identified. As we will see
below, the matrix elements $\Gamma_{lm}$ are time dependent, implying a 
non-Markovian evolution, causing a gaussian decay of the polarization
for small times. Also, the polarization will show temporal decay and growth
periodically, displaying the underlying Hamiltonian evolution of the spin-1/2
and the bath taken together. Let us start by rewriting our general solution for
the polarization of the spin-1/2 particle as a function of time given in
Eq.\ref{manyspinpol} as 
\begin{equation}
 \hat P_A(t) - \hat g (t) = M \hat P_A(0),
\end{equation} 
where $\hat P_A, \hat g$ are the column vectors of $\vec P_A,\vec g$
respectively, and $M$ is a 3 $\times$ 3 matrix given as
\begin{eqnarray}
M = \left( \begin{array}{ccc} 
f+\tilde\Pi_{xx}  & h_z +\tilde\Pi_{xy}& -h_y+\tilde\Pi_{xz}\\
-h_z+\tilde\Pi_{xy}  & f +\tilde\Pi_{yy}& h_x+\tilde\Pi_{yz}\\
h_y+\tilde\Pi_{xz}  &- h_x +\tilde\Pi_{yz}& f+\tilde\Pi_{zz}
\end{array}\right).
\end{eqnarray}
Then the equation of motion for the polarization vector can be written as
\begin{equation}
{d\hat P_A (t) \over dt} = D \hat P_A(t) +\hat R(t),
\end{equation}
where the matrix $D= \frac{dM}{dt} M^{-1}$ and the inhomogeneous term is given by
$\hat R(t) = d\hat g(t)/dt - D \hat g(t)$. Now, a straightforward comparision
of this with Eq.\ref{eqmo} yields expressions for the matrix
elements $\Gamma_{lm}$ and the effective magnetic field $\vec B_{eff}$ (in
units of $\hbar/\mu_B$) as
\begin{equation}
\Gamma_{lm}= D_{lm} + D_{ml} - \delta_{lm} Tr D - i\sum_n \epsilon_{lmn}
\hat R_n,
\end{equation}
and
\begin{equation}
\vec B_{eff}.\hat e_l = {1\over 2} \sum_{mn}\epsilon_{lmn}D_{mn}.
\end{equation}
It is easy to check at small times , that $\Gamma_{lm} \sim \alpha t$ and $|\vec
B_{eff}| \sim \beta(1-\delta t^2)$.

In the above equation for the polarizaion vector, the time developement of the
different components are coupled. Let us rewrite the matrix $D$ interms of the
symmetric part $D_s$ and the antisymmetric part $D_a$ as $D=D_s+D_a$. The
antisymmetric part gives the effective magnetic field, which causes a 
standard precession of the polarization, with no decay or decoherence.
The symmetric part
causes a decay or growth of the polarization components.
We can change to the diagonal basis of the
matrix $D_s$, and the transformed vectors are $\tilde P, \tilde R$. The time
developement of the polarization components can be found directly by
integration, and $\gamma_i$ (eigenvalue of $D_s$) determines
the decay or growth of the polarization component $\tilde P_i$. For short times,
we have $-\gamma_i\sim t$, and the polarization is given by 
\begin{equation}
P^2(t)\approx \sum_i \tilde P_i^2(0) {\rm e}^{2 \int_0^t \gamma_i dt},
\end{equation}
which implies a gaussian decay.
If all the elements of the matrix $\Gamma$ (or $D$) are independent of time,
implying time-independent decay cosntants $\gamma_i$, then we would have an
exponential decay which is the hallmark of a markovian evolution.

For an illustration, let us consider a purely vector-polarized bath,
where the density matrix in each sector of the bath spin $I_B$ is given by 
$\rho_{I_B} = [\mathcal{I} +
{1\over {I_B}} P^z_{I_B} I^z_B]/(2I_B+1)$.
Here, there is no tensor polarization, and $\vec{h}(t) = h(t)
\hat{z}$ and $\vec{g}(t) = g(t)\hat{z}$.
The expressions for the effective magnetic field and the $\Gamma$ matrix
can be directly calculated for this case,
\begin{eqnarray}
\vec{B}_{eff} = \frac{f\dot{h} - \dot{f}h}{f^2 + h^2} \hat{z}
\end{eqnarray}
\begin{eqnarray}
\Gamma = \left( \begin{array}{ccc}
 -{d\over dt}\log{f}  & 
-i f {d\over dt}( {g\over f}) & 0 \\ i f {d\over dt}( {g\over f}) 
 & -{d\over dt} \log{f}  & 0 \\ 0 & 0 & {d\over dt}\log{f\over {f^2 + h^2}} 
\end{array}\right).
\end{eqnarray}
\begin{figure}[htb]
  \begin{center}
    \includegraphics[width=8.0cm]{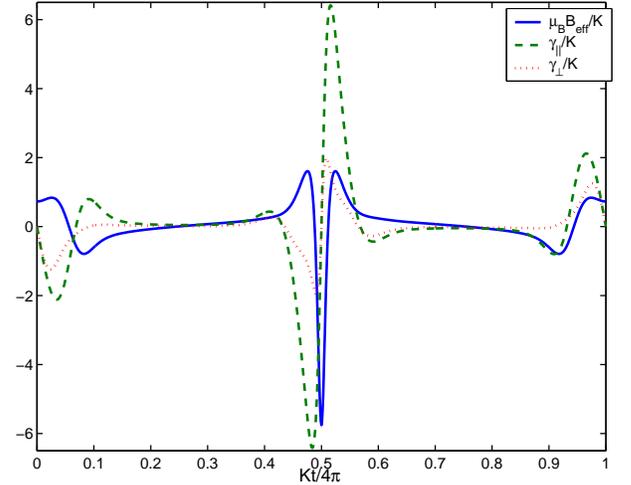}
     \end{center}
\caption{The decay functions and the effective magnetic field
are plotted as a function of time when the initial states of the qubit and bath
are
$\rho_A(0) = \vert \downarrow \rangle \langle \downarrow \vert$ and
$\rho_{B}(0) = \sum_I \lambda_I/(2I+1) (\mathcal{I} + \vec{P}_I.\hat{z})$
respectively.
The bath is composed of $N = 100$ particles with the bath spin distribution
given by $\lambda_I = \exp(-0.1I^2)$. Note that, with $\hbar=1$,
both the decay functions $\gamma_\parallel$ and $\gamma_\perp$ have a
dimension of energy.}
\end{figure}
The diagonal elements of the above matrix are related to the decay functions
through, $\Gamma_{ii}= 2\gamma_i - Tr D$. We have,
\begin{equation}
\gamma_1 = \gamma_2 \equiv\gamma_\perp  = {d\over dt}\log f,
\end{equation}
\begin{equation}
\gamma_3 \equiv \gamma_\parallel = {1\over 2}{d\over dt} \log ({f^2+h^2})
\end{equation}
 We see that  the polarization of the bath does produce an effective
 magnetic field $B_{eff}$ which is time dependent, and causes a 
precession of $\vec{P}_A(t)$. For short times, the effective magnetic field
is $B_{eff}\approx K/\mu_B$ which is about a few gauss using a typical
interaction strength $K\sim 10^{-8} {\rm eV}$ in QDQC systems. In Fig.6 we have plotted the effective magnetic
field, and the decay/growth functions with time, for a representative
gaussian bath-spin distribution. 
As can be seen from the figure, the effective magnetic field and $\gamma_i$
show a nonmonotonic behavior, and they take both positive and negative values.
A negative $\gamma_i$ implies a decay of the
corresponding polarization components, and a positive $\gamma_i$ implies a
growth of polarization. Though $\gamma_i$ are time dependent, they
satisfy the condition $2\gamma_{\perp}>\gamma_{\parallel}$, consistent
with the general constraint found by Gorini et al\cite{gorini}.

Finally, we determine the effective magnetic field when the initial bath state
is given by  $\rho_{B}(0) = \vert \uparrow \uparrow ........\uparrow
\rangle \langle \uparrow \uparrow ........\uparrow \vert$, an example which we
have discussed in section IV.B. Here the bath is in a pure state, and the
tensor polarization is nonzero as discussed in the earlier section. In this
case, we have
\begin{equation}
\vec{B}_{eff} = \lbrace 1-\frac{2}{N+1}\sin^2(N+1)Kt/4 \rbrace \frac{KN}{2}\hat{z},
\end{equation}
\begin{equation}
\gamma_\parallel=2\gamma_{\perp} = -\frac{NK}{(N+1)}\sin(N+1)Kt/2.
\end{equation}
Here the strength of the effective magnetic field is about a few hundred
gauss for $N=100$, $K\sim 10^{-8} {\rm eV}$.

\section{Conclusions}
The interaction of a central spin-1/2 particle with the nuclear spins 
in quantum dots is modeled by a Heisenberg exchange interaction. The dynamics
of initial direct product states of the qubit and the nuclear spin bath are
investigated. The decoherence of the spin-1/2 particle can be seen from the
the decay of its polarization through the Hamiltonian time evolution.
The qubit polarization as function of time is explicitly calculated for
any nuclear bath with a spin-conserving internal dynamics, for any bath-spin
distribution and any polarizations. The time developement of the qubit 
polarization has nonmarkovian features.
The qubit polarization shows a gaussian decay/decoherence for short times.
For a typical interaction strength of the qubit and the nuclear spins,
$K\sim 10^{-9}{\rm eV},$ the decoherence time scale is about 100 nanoseconds.
 For longer times, the
polarization shows nonmonotonic behaviour, eventually displaying a periodicity
in time.
The gaussian decay time scale depends on the bath-spin distribution, a larger
width of the distribution leading to smaller time scales. The vector and tensor
spin polarizations that may be present in the nuclear spin bath have a tendency
to increase the gaussain decay time scale.

\end{document}